\documentclass[krantz1,ChapterTOCs]{krantz} 
\usepackage{fixltx2e,fix-cm}
\usepackage{amssymb}
\usepackage{amsmath}
\usepackage{graphicx}
\usepackage{subfigure}
\usepackage{makeidx}
\usepackage{multicol}
\usepackage{color}
\usepackage{chicago}

\begin{document}


\title{Handbook of ABC} 
\author{Yours Truly}
\maketitle


\mainmatter

%
%
%
\chapter{ABC in Nuclear Imaging}

 \contributor{Y. Fan$^1$, S. R. Meikle $^2$, G. Angelis$^2$, A. Sitek$^3$}{$^1$ School of Mathematics and Statistics, UNSW, Australia}{$^2$ Medical Radiation Sciences, University of Sydney, Australia}\\{$^3$ Philips Research North America, Cambridge, MA, USA}

\section{Introduction to Nuclear Imaging}\label{intro}

{Nuclear imaging} technologies produce noninvasive measures of a broad range of physiological functions,  using externally detected electromagnetic radiation originating 
from radiopharmaceuticals administered to the subject. The main nuclear imaging modalities PET ({positron emission tomography}) and SPECT (single photon emission computed tomography) are the backbones of the field of molecular imaging, where they are used extensively in both clinical settings and pre-clinical research with animals to
study disease mechanisms and test effectiveness of new therapies.  Typical applications include glucose metabolism studies for cancer detection and evaluation, and cardiac imaging; imaging of blood flow and volume; and it is one of the few methods available to neuroscientists to non-invasively study biochemical processes within the living brain such as receptor binding, drug occupancy, or neurotransmitter release.

\begin{figure}
\centering
\includegraphics[width=\textwidth]{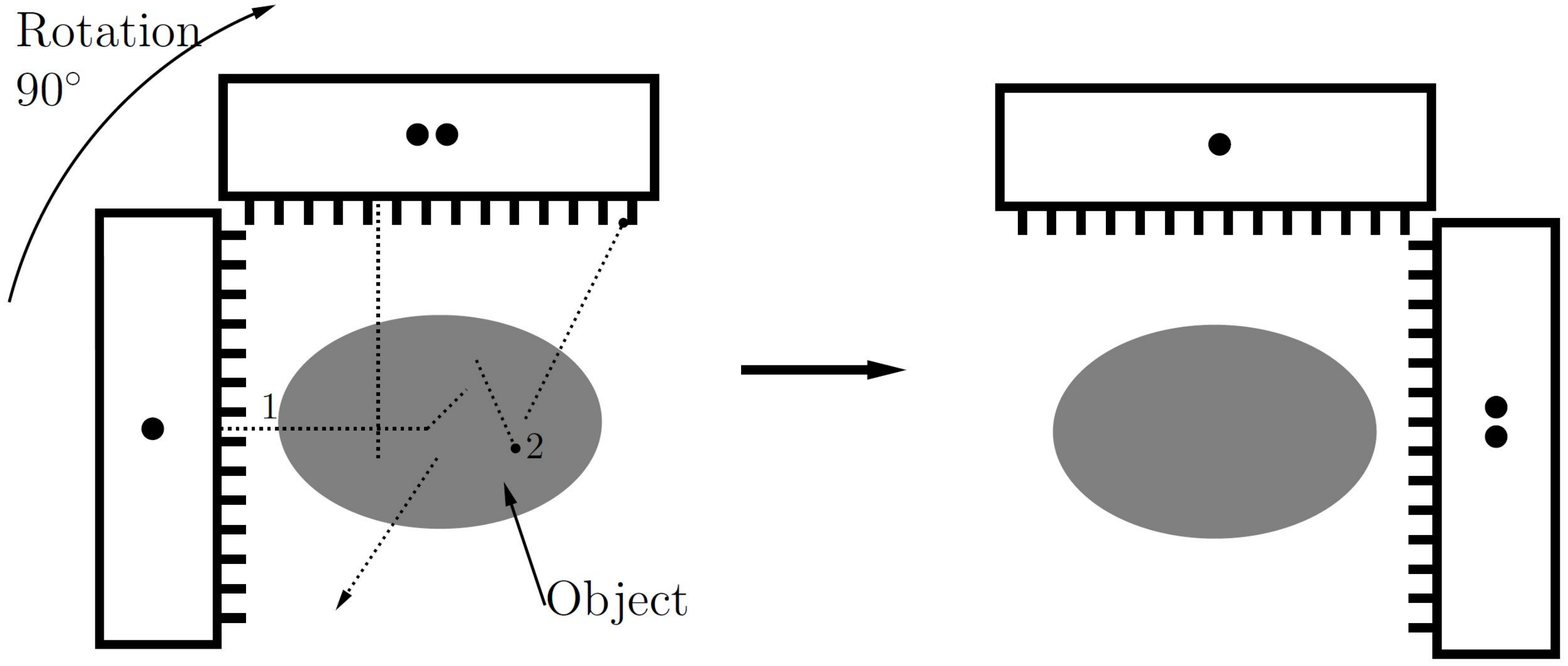}
\caption{\it  Data acquisition in SPECT using a dual head scanner. Each of two
heads is an independent gamma camera (Anger, 1964). 
With this acquisition setup the system
needs only 90 degree rotation to acquire counts from all directions in a 2D plane
around the object. {There are many intermediate steps (in the order of 64) between configuration on the left and right at which the data is acquired.} Dotted lines illustrate hypothetical paths of gamma photons
which can be absorbed by the object (attenuated) (\#2) or scattered and then detected
(\#1). These are two examples of many possible interactions. Reproduced from Sitek (2014).} 
\label{fig:SPECT}
\end{figure}
\begin{figure}
\centering
\includegraphics[width=\textwidth]{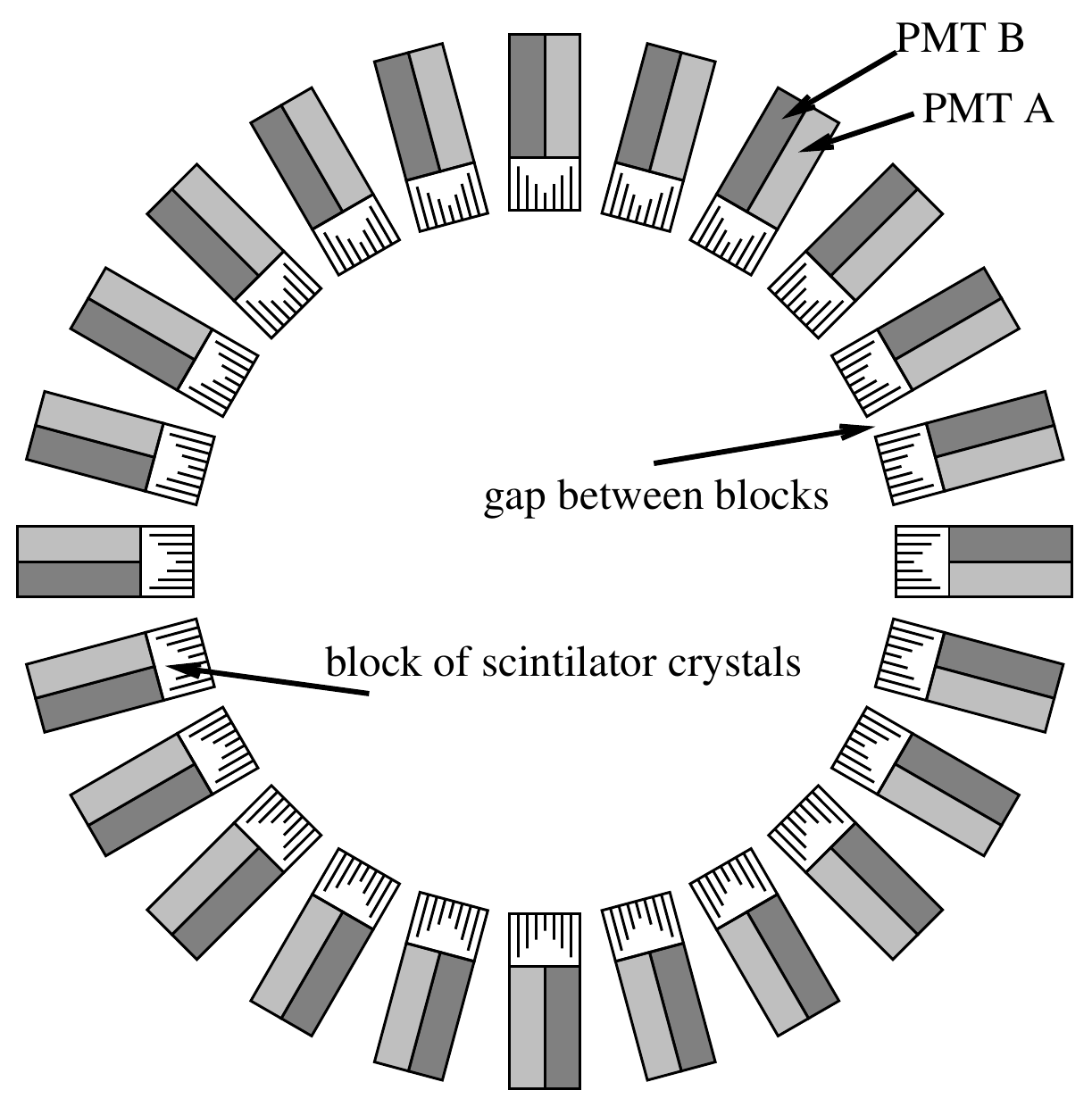}
\caption{\it PET camera comprising 24 PET block detectors. Photomultiplier tubes (PMTs) are attached to scintillation crystals for signal amplification and interaction localization.  
{Two of the four PMTs per detector block needed to localize a gamma ray interaction are shown as PMT A and PMT B.} Reproduced from Sitek (2014). }
\label{fig:PET}
\end{figure}

{In nuclear imaging, a subject is administered with a small amount of radiopharmaceutical called a {\it tracer}. Radiation is created when the nuclei of the tracer decay and produce photons in the range 35-511 keV that are then detected by radiation sensitive detectors external to the subject. The energy of the photons must be high enough to allow the photon to leave the subject's body but low enough to allow absorption in the detector.  Photons can be produced either as a direct product of the nuclear reaction that occurred, or indirectly. For example,  SPECT is based on single photon detection produced by the decay of the radioisotopes. {A gamma camera} \cite{anger64} is rotated around the subject and acquires photon counts at different projection angles, typically a full 180 degree set of projections are needed. {Figure \ref{fig:SPECT} shows} two of the positions during data acquisition.
On the contrary, PET is based on indirect photon detection produced by positron annihilation where a radioactive decay produces a positron which annihilates with an electron and produces a pair of  photons travelling in opposite directions along a straight line path. The photons are detected by a large number of detectors surrounding the subject, forming a ring. The PET detector ring is stationary, see Figure \ref{fig:PET}. Basic physical principles underlying PET and SPECT imaging and instrumentation for data acquisition can be found in the reviews of \shortciteN{cherry04}, \shortciteN{wernick04}.}

The goal of imaging is to study the  concentrations of radioactive nuclei that are in the imaged object assuming that they are attached to the tracer molecules. The description is simplified by subdividing the imaged volume into non-overlapping equal sized volume elements called {\it voxels}, and only the numbers of nuclei inside each voxel are considered.  In a typical setup of the inverse problem,
the quantity of interest is the expected number of decays per voxel which leads to the data being modelled as Poisson distribution.  Nuclear imaging can be extended to so called {\em dynamic imaging} where changes in distribution of the tracer are investigated over time.  This is done by dividing the time in which the changes are observed (typically in the order of 30 minutes to an hour) into time frames (about 20 to 60) and reconstruct voxelized time frames independently. The reconstructed time frames are then analyzed by algorithms such as the one presented in this chapter.  

The voxelized {\it image} data are reconstructed from acquired data (counts) by any one of a number of reconstruction algorithms, see for example \shortciteN{qil06}. This {reconstruction} is called the {\em tomographic reconstruction}. The primary factors limiting reconstructed image quality are detector resolution, which determines the maximum resolution of the reconstructed images; and the total number of detected counts, which determines the minimum noise level that can be achieved at the maximum resolution. {In the following discussion, we restrict our attention to model-based reconstruction algorithms, which are of more interest to the statistical community.}
In model-based reconstruction approaches, a probabilistic model is used to account for the physical and geometric factors that affect photon detection. In its simplest form, the image reconstruction can be seen as a problem of parameter estimation, where the acquired data (counts) are Poisson random variables with mean equal to a linear transformation of the parameters. Let ${\bf y}$ be the measured projection data, and ${\bf x}$ be the unknown image; ${\bf x}$ is related to ${\bf y}$ via 
$$E({\bf y})={\bf P}{\bf x}.$$
The projection matrix ${\bf P}$ models the probability of an emission from each voxel element in the source image being detected at each detector element. In simple terms,  ${\bf x}$ is the reconstructed image, where each element of the matrix ${\bf x}$ corresponds to a voxel element in the image. ${\bf y}$ is the measured count, assumed to have Poisson distribution.  To give an idea of the scale of the problem, a single 3-dimensional scan could produce $10^7-10^8$ counts, with $10^6$ image parameters to be estimated. Many methods have been proposed for solving the inverse problem, starting with the EM algorithm of \shortciteN{sheppv82}, to the ordered subsets algorithm of \shortciteN{hudsonl94}, to many more sophisticated algorithms which deal with the problem of ill-conditioning often arising in PET applications
 (where the solutions to the inverse problem are sensitive to small changes in the data). \shortciteN{fessler96}, \shortciteN{leahyq00} and \shortciteN{qil06} provide detailed reviews on the statistical challenges in model-based reconstruction methods. \shortciteN{sitek14} discusses in depth the statistics of detected counts.

The 3-dimensional images of the radio tracer distributions when monitored over time, provide insights into physiological 
state of the organism in vivo. This dynamic imaging is often referred to as {\it functional imaging}. {Functional imaging} focuses on how tracers accumulate and clear from the tissue, enabling the physiological  function associated with that tissue to be measured. Typically the changes are characterised
by using models of biological processes occurring in the voxel or the region of interest (ROI), which is a group of voxels corresponding to a particular anatomical region. As explained above, dynamic data are obtained by dividing the total acquisition time into intervals or {\em time frames}, and the data acquired in each time frame are reconstructed independently, representing the average concentration of the tracer in a voxel over the time interval. It is possible to process the data so that the correlation between time frames is taken into account, but in practice simpler approaches are used because of the ease of processing. Analyses of the 4-dimensional spatio-temporal data set 
often proceeds by modelling the changes in concentration of the tracer using appropriate compartmental models of temporal data at each voxel or averaged groups of voxels (ROIs). These temporal data are termed the {time activity curve} (TAC). 
Compartmental models provide estimates of biologically meaningful parameters. The parameters of the models can be estimated for each voxel separately (as opposed to a group of voxels, ROI) to produce parametric images (one 3D image for each parameter) that describe important physiological information about the subject.
An important consideration in  parametric image estimation is 
robustness to noise, as noise in the voxel TAC can be high.  Additionally, any estimation  has to be very fast due to the
large number of voxels associated with each image and large numbers of TACs to process.

In this chapter, we first briefly describe compartmental models in PET in Section \ref{sec:compart}, we then introduce a simple ABC 
algorithm in the context of PET kinetic modelling in Section \ref{sec:abc}. Section \ref{sec:example} provides a detailed example
of ABC implementation for a neurotransmitter response model and in Section \ref{sec:discuss}, we conclude with some discussions about the potential of ABC in medical imaging.

\section{Compartmental models in PET}\label{sec:compart}

As discussed, PET is a {functional imaging} technique so that given a time sequence of images, one can monitor the 
interaction of a particular radiotracer molecule with the body's physiological processes. For instance, blood flow can be measured
by using radioactive water (with $^{15}$O {replacing  $^{16}$O in water H$_2$$^{16}$O molecules, by bombarding them with protons}) as a tracer and metabolism can be measured with a radioactive glucose analog.

{Kinetic models} for PET typically derive from the one-, two-, or three-compartment model with a model input function. In PET, one normally assumes that all tissues in the body see the same input function and this is typically a measured concentration of radioactivity  in the blood plasma during the experiment.   In compartmental modelling, it is assumed that within a voxel, whatever radioactive species contribute to the radioactive signal are in uniform concentration and can be characterised as being in one or more unique states. {Assuming the system is in steady state, } each of these states is assigned a compartment, which in turn is described by {the rates of a change in concentration within a} single ordinary differential equation. The coefficients of the differential equations or the kinetic parameters are reflective of inherent properties of the particular radiotracer  molecule in the system, providing information about any hypothesised processes. 

\begin{figure}
\centering
\includegraphics[width=\textwidth, height=240pt]{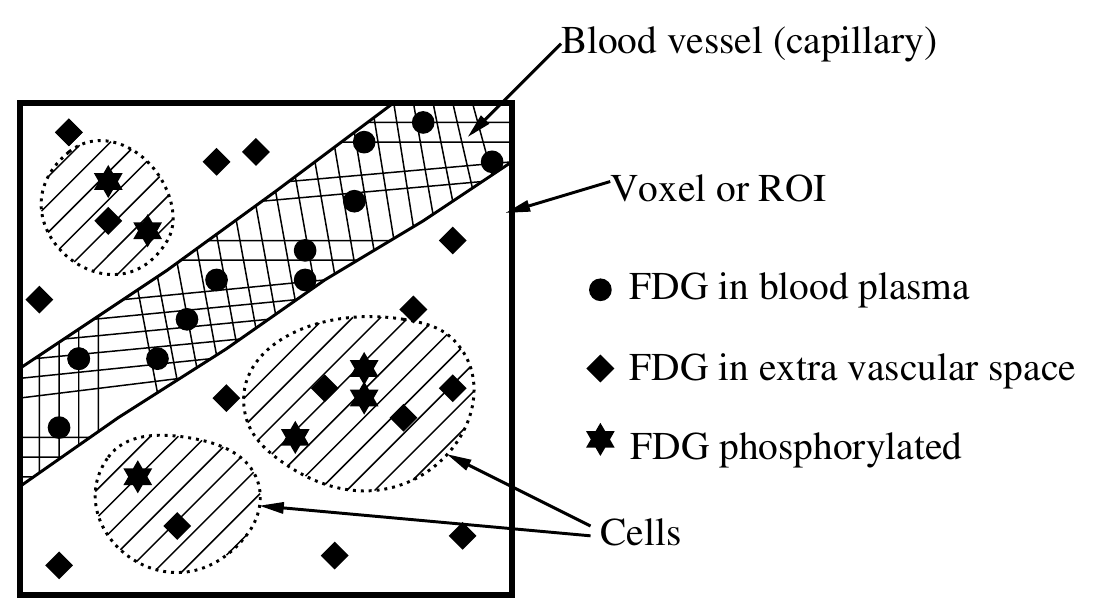}
\caption{\it  Representation of the voxel or ROI. {The tracer (in this example FDG) is assumed to be in either of three states: in blood plasma, in the extra
vascular space, or in a phosphorylated state within the cell.} Reproduced from Sitek (2014).} 
\label{fig:FDGcomps}
\end{figure}

As an illustration of the {compartmental model}, consider the example given in \citeN{sitek14}, Chapter 5. Figure \ref{fig:FDGcomps} illustrates the possible physiological states of  
the tracer compound $^{18}$FDG, a glucose analog. The compound is delivered to the blood, and transported into the cells. Three possible states can be identified: (1)$^{18}F$-Fluorodeoxyglucose ($^{18}$FDG, analog of glucose) within the plasma, (2) unmetabolized $^{18}$FDG
 present in the cells or the interstitial spaces between cells, and (3) phosphorylated $^{18}$FDG which is trapped in the cell (Figure~\ref{fig:FDGcomps}). Compartmental models are then built by describing the connections between the states of the molecules, describing the influx to, and efflux from each compartment,  in the form of ordinary differential equations.

\begin{figure}
\centering
\includegraphics[width=\textwidth, height=240pt]{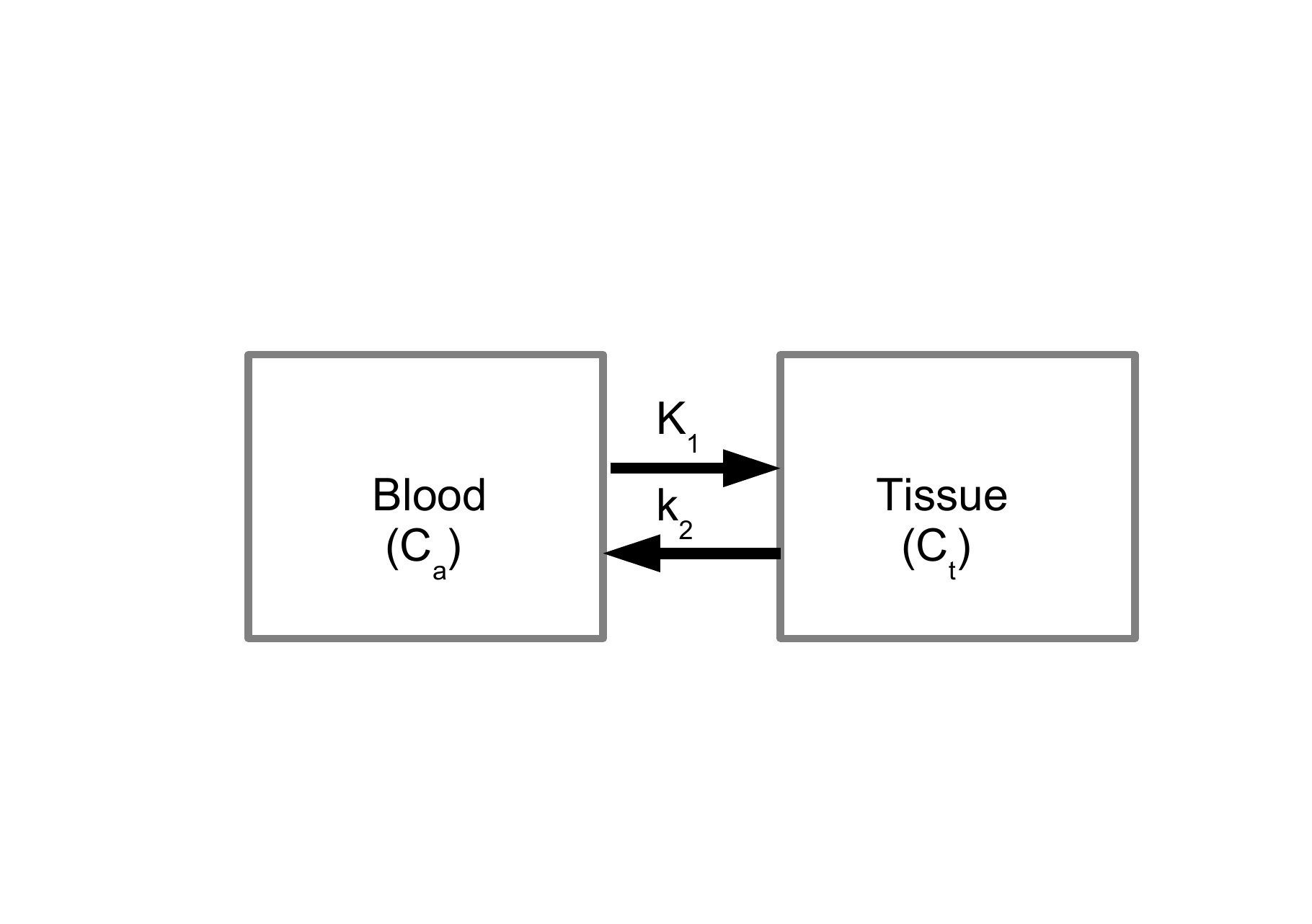}
\caption{\it  One-tissue compartment model describing the flow of the tracer between blood ($C_a$) and tissue ($C_t$). {$K_1$ and $k_2$ are the kinetic rate constants,  see Equation \ref{eq:onecompart}.}}
\label{fig:1tcm}
\end{figure}

The one-tissue {compartmental model} is the simplest model that frequently arises in PET applications, describing the bidirectional flux of tracer between blood and tissue. See Figure \ref{fig:1tcm} for a pictorial depiction of the model.  The one-tissue compartment model
is characterised by the tracer concentration in the tissue over time $C_t(t)$, the arterial blood (or blood plasma input function) $C_a(t)$ and two first-order kinetic rate constants $(K_1, k_2)$. The tracer flux from blood to tissue is $K_1C_a(t)$ and the flux from tissue to 
blood is $k_2C_t(t)$, so the net tracer flux into tissue is given by the ordinary differential equation as
$$\frac{dC_t(t)}{dt}=K_1C_a(t)-k_2C_t(t)$$
which is solved to obtain
\begin{equation}\label{eq:onecompart}
C_t(t)=K_1C_a(t) \otimes  \exp(-k_2t)
\end{equation}
where the symbol $\otimes$ denotes the one-dimensional convolution. For a PET image, $C_t(t)$ is the measured radioactivity 
concentration in a voxel or ROI, $C_a(t)$ is the arterial blood concentration of {the} tracer measured in a sample drawn during a scan. If the PET data
are not corrected for physical decay, the parameter $k_2$ includes a component of radioactive decay. For further interpretation of kinetic rate parameters, see \shortciteN{morris04}.

More complex compartmental models distinguish different biochemical or physiological states of the tracer in tissue. After entering a cell, the tracer is available for 
binding in a free form at the concentration $C_1(t)$. Free tracer can specifically be bound to its target molecule, {with concentration} $C_2(t)$, but it may also specifically bind to some cell components that are not known in detail, $C_3(t)$. The system of differential equations can be derived analogously to the one-tissue compartment model, but {is} much more complex, with six unknown parameters that {may be} difficult to estimate. In practice the system is often reduced to a two-compartment model by treating free and non-specifically bound tracer as a single compartment provided that the rates of exchange between the free and bound states are sufficiently rapid compared with the net influx into the combined compartment. Authoritative reviews on the subject can be found in  \shortciteN{morris04}, \shortciteN{innis07} and \shortciteN{gunnssp15}.

\section{Parameter estimation in {compartmental models}}
The amount of data available to fit the model is relatively small, typically around 20-40 time points per voxel or ROI. The estimation of parameters based on these data, sometimes 10 or more of them, can be non-trivial. In more realistic and complex models, parameter identifiability becomes an issue due to the sparsity of data. Therefore the adoption of a particular model is by necessity a simplification of the truth  (\shortciteNP{gunngtac02}).  Robustness of parameter estimation in the presence of high level of noise, particularly in voxel-wise estimations, where the noise-to-signal ratio can be high, poses another difficulty. In addition, since a separate estimation procedure has to be performed for each voxel, this might typically be around one million voxels, computational speed needs to be taken into consideration.

{A typical} approach to parameter estimation in kinetic modelling proceeds via a variety of least squares fitting procedures (\shortciteNP{carson86}, \shortciteNP{feng96}); weighted integration (\shortciteNP{carsonh86}) or basis function techniques (\shortciteNP{gunn97}). Many authors have commented on the difficulties with using non-linear least squares methods, particularly with noisy data, often failing to converge, producing estimates with large variances {(which can be the case even in noiseless data) } (\shortciteNP{gunngtac02}, \shortciteNP{alpert09}). This has led to methods that employ penalised optimisation to stablise parameter estimates (\shortciteNP{zhou01}, \shortciteNP{gunngtac02}).  

Whilst the limitations of the basis function technique of \shortciteN{gunn97} are well known, its simplicity and ease of implementation has made it a preferred method for parameter estimation of kinetic models for PET data. The basic  idea is to linearise the kinetic equation, and then use (weighted) least squares methods to obtain parameter estimates. Consider for example the one-tissue compartment model (\ref{eq:onecompart}), the parameter $K_1$ is linear whilst the parameter $k_2$ is non-linear. The non-linear term is then dealt with by choosing a discrete spectrum of parameter values for $k_2$ and forming the corresponding basis functions
$$ B_i(t)=C_a(t)\otimes \exp (-k^i_2 t)$$
for $i=1,\ldots, n$, where the values of $k^i_2$ are taken from a physiologically plausible range of values for $k_2$. Equation (\ref{eq:onecompart}) then becomes linear in $K_1$, where
$$ C_t^i(t)=K^i_1B_i(t).$$
The parameters $K^i_1$ can now be solved for each basis function $B_i(t)$ using  linear least squares, and the parameter set $(K_1^i, k_2^i)$ that produces the minimum residual sum of squares is taken as the optimal solution, (\shortciteNP{cunninghamj93}, \shortciteNP{meikle98}).  \shortciteN{gunn97} reported that in their experimentation, only 100 basis functions were needed to obtain good results, making the method very time efficient.
It is interesting to note that the idea of fitting a spectrum of values of $k^i_2$ and then choosing the most likely value according to some goodness of fit criterion,  is very similar to ABC, where ABC formalises the selection of the candidate parameter set with a prior distribution. Whilst the method of  \shortciteN{gunn97} is not formally Bayesian, the authors note the superior performance of the estimation when a constraint or a bounded region is placed on the non-linear parameters, thus  implicitly placing a prior distribution on the unknown parameters.

{The scarcity of data in kinetic modelling lends itself naturally to Bayesian modelling, where inclusion of priors can provide better estimates. }
This approach has been advocated more recently by several authors (\shortciteNP{zhouaj13}, \shortciteNP{alpert09}, \shortciteNP{malaves15}).  Most applications of Bayesian modelling in medical imaging proceed in a frequentist fashion, that is, one often simply finds the maximum a posteriori (MAP) estimate of the posterior using any number of optimisation tools, see for example \shortciteN{lin2014}. 
Recently, \shortciteN{malaves15}, \shortciteN{sitek14}  has advocated a proper treatment of Bayesian inference in the medical imaging community, given that 
uncertainty quantification is particularly relevant when the observational data has a very low signal to noise ratio. 

Typically, the full Bayesian inference proceeds by assuming an error model for the time activity curve. The most common model is the independent Gaussian error model, with the variance at each time point assumed to be proportional to the observed data point. Markov chain Monte Carlo (MCMC) is the default posterior sampling method. However despite its wide usage, the Gaussian error model is often not appropriate. \shortciteN{zhouaj13} found that a $t$-distribution worked better for the examples they studied. In reality, the error distribution is highly positively skewed at time points with low activity if {a} non-negativity constraint is used with reconstruction, and more symmetric at higher activity time points. In simulation studies,  Poisson error is often {introduced} to the deterministic data.  A second difficulty is that MCMC itself requires tuning and convergence assessment. {While the former can be automated to some extent by automatic tuning algorithms (\shortciteNP{garthwaitefs15}), the latter would ideally require repeat analyses at dispersed starting points. This can be computationally infeasible when the analyses involves hundreds of thousands of repeat simulations.}

\section{A  simple ABC algorithm for kinetic models}\label{sec:abc}
ABC offers an alternative to MCMC. Traditionally, ABC is used when the likelihood function is not tractable. In the current setting, ABC offers a way of computing full Bayesian analyses without the need to specify an exact error distribution: we only require the ability to simulate summary statistics. 
The most obvious advantage is its ease of interpretation and application, which makes fully Bayesian inference easily achievable for practical users of Bayesian methodology. In this chapter, we will
restrict our attention to the simplest of ABC algorithms, the standard rejection sampling method. For the parameter vector {${\boldsymbol \theta} = (\theta_1, \ldots, \theta_p) '$}, this is achieved by the following three steps:
\begin{enumerate}
\item {S}ample parameters {$\theta_i, i=1,\ldots,p$} from the {sampling} distribution, Uniform$(a_i,b_i)$
\item {C}ompute $\hat{C}_t(t)$ using ${\boldsymbol \theta}$, and the corresponding $S_{sim}$
\item {R}etain ${\boldsymbol \theta}$ if $\sum_t|S^t_{sim} - S^t_{obs}| <\epsilon$
\end{enumerate}
The sampling distributions Uniform$(a_i,b_i)$ are proportional to the prior distributions for each parameter, Uniform$(a^*_i,b^*_i)$, we will discuss how to obtain a
good sampling distribution in Section \ref{sec:sampling}. $\hat{C}_t(t)$ is the estimated activity concentration, using the trial {value} of ${\boldsymbol \theta}$. For example,  ${\boldsymbol \theta}=(K_1,k_2)$ if using Equation (\ref{eq:onecompart});  $S^t_{sim}$ and $S^t_{obs}$ are the simulated and observed summary statistics, respectively,  at the  {$t$-th} time point. $\epsilon$ is a predetermined error tolerance value. {The choice of summary statistics will be discussed in Section \ref{sec:example}.}

It is clear from the above, that in repeated estimations for different voxels, Steps 1 and 2 do not need to be repeated.  {This is because the values $\hat{C}_t(t)$ computed for one voxel can be reused for others and the additional computational cost in Step 3 is relatively small. }

In the algorithm above, we have not replicated the noise in the data. Since we are not interested in estimating the parameters in the error distribution, those are {considered} nuisance parameters. What we assume here is that there exist summary statistics that are (nearly) sufficient for the kinetic parameters. We will discuss the selection of summary statistics in more detail in the example section.  

This simple form of ABC is similar to the popular basis function approach of \shortciteN{gunn97}, where the {summary statistics are} just taken as the original data. ABC {formalises} the constraints on the parameters in the form of a prior, and instead of using least squares for some of the parameters, ABC samples all parameters.
{In addition, the ABC method provides parameter uncertainty estimation by probabilistically retaining some of the sampled parameters.}

\section{Application to a neurotransmitter response model}\label{sec:example}

Development of neurochemical assays that capture temporal signatures is critical because the neurotransmitter dynamics
may encode both normal and abnormal cognitive or behavioural functions in the brain. The elucidation of specific patterns of 
neurotransmitter fluctuations are beneficial to the study of a wide range of neuropsychiatric diseases, including alcohol and substance abuse disorders (\shortciteNP{morris05}, \shortciteNP{normandin12}).

 \shortciteN{morris05} developed a new model, called ntPET, for  quantifying time-varying neurotransmitter concentrations. The new model  enhances the standard tracer kinetic model, accounting for both time-varying dynamics of the radiotracer $[^{11}$C]raclopride and 
the endogenous neurotransmitter dopamine that competes with it for the same D2 receptor binding sites. For the input function, a reference region approach is used instead of arterial sampling, {where the activity concentration measurements in the reference region of tissue are assumed to contain negligible specific binding signal
(\shortciteNP{morris04}).} Experimental data are acquired in two separate PET scans, one conducted with the subject at rest and the
other immediately following a stimulus. \shortciteN{normandin12} further developed this model to be used with a single scan session and 
proposed a basis function approach for the simplification of computation; they call the method lp-ntPET (linear parametric-neurotransmitter PET). In our simulation studies, we will generate simulated data using ntPET, and fit the model lp-ntPET to the simulated data, since the latter is a simplification of the former.

The operational equation for the  lp-ntPET model takes the form
\begin{equation}\label{eq:lpntPET}
C_t(t)=R_1C_R(t) + k_2\int_0^t C_R(u)du - k_{2a}\int_0^t C_t(u)du-\gamma\int_0^tC_t(u)h(u)du
\end{equation}
where $C_t(t)$ and $C_R(t) $ are the concentration of the tracer in the target tissue and reference regions, respectively. The parameters $R_1$, $k_2$ 
and $k_{2a}$ describe the kinetics of tracer uptake and retention in the tissue. {The parameter $\gamma$ describes} the neurotransmitter response magnitude.

{The function $h(t)$ describes the non-steady state component of the kinetic  model }(with $\gamma$ encoding the magnitude), given by
$$ h(t)=\left(\frac{t-t_D}{t_P-t_D}\right)^{\alpha}\exp\left(\alpha\left[1-\frac{t-t_D}{t_P-t_D}\right]\right)u(t-t_D)$$
where $u(t)$ is the unit step function.  The variable $t_D$ is the delay time at which the response starts relative to {the} start of scan, $t_P$ is the peak time of maximal response magnitude, $\alpha$ is the sharpness of the function. The lp-ntPET model has seven parameters, four that describe tracer kinetics and response magnitude 
 $(R_1, k_2, k_{2a}, \gamma)$, and three describing the time course of the neurotransmitter/activation response $(t_D, t_P, \alpha)$. This formulation is a simplification of the ntPET model which has eleven parameters.

Equation (\ref{eq:lpntPET}) can be expressed in matrix form $y=Ax$,  as
\begin{equation}\label{eq:ls}
\left [ 
\begin{array}{c}
C_t(t_1)\\
\vdots\\
C_t(t_m) 
\end{array}
\right] 
=
\left [ 
\begin{array}{cccc}
C_R(t_1) &\int_0^{t_1}C_R(u)du &-\int_0^{t_1}C_t(u)du &-\int_0^{t_1}C_t(u)h(u)du\\
\vdots &\vdots &\vdots &\vdots\\
C_R(t_m) &\int_0^{t_m}C_R(u)du &-\int_0^{t_m}C_t(u)du &-\int_0^{t_m}C_t(u)h(u)du
\end{array}
\right] 
\times
\left [ 
\begin{array}{c}
R_1\\
k_2\\
k_{2a}\\
\gamma 
\end{array}
\right] .
\end{equation}
So, for fixed values of $t_P, t_D$ and $\alpha$ in the function $h(t)$, the above representation can be solved using linear least squares. 

\shortciteN{normandin12} {propose} an efficient computational algorithm for parameter estimation {for} lp-ntPET. The idea is  similar to the basis 
function method of \shortciteN{gunn97}. Setting the basis function to be $B_i(t)= \int_0^t C_t(u)h_i(u)du$ (this corresponds to the last column entry of the matrix $A$),  then for basis function $B_i(t)$, a {weighted least squares} solution is obtained for Equation (\ref{eq:ls}), where $\hat{x}=(A^TWA)^{-1}A^TWy$ with the weight matrix having diagonal elements inversely proportional to the variance of the PET measurement of $C_t$ in the matching row of the matrix equation, since it is commonly assumed that the variance of the tracer concentration is proportional to the observed value. A similar assumption is made in most Bayesian models using Gaussian error assumption, see for example \shortciteN{zhouaj13}.  Clearly, in the presence of high noise, such an assumption can lead to poor parameter estimation. Finally, a large library of basis functions are calculated over different combinations of $t_D, t_P$ and $\alpha$, and the parameter set that minimises the residual sum of squares is then chosen as the final estimate. If the non-negativity constraint is to be used, for example, for the parameter $\gamma$, then an iterative weighted least squares approach is adopted. 

In the next section, we consider the application of ABC to the problem of neurotransmitter response modelling described above. We obtain simulated data, using the nt-PET model, and use ABC to fit {the simpler lp-ntPET model} to the data at varying levels of noise. The noise is  Poisson with a mean proportional to the simulated activity concentration. Simulation data are obtained over 60 time frames each with one minute duration.

\subsection{{Prior and sampling distributions}}\label{sec:sampling}
{For simplicity, we use the Uniform distributions $U(a_i^*, b_i^*), i=1,\ldots, 7$ as the prior distributions for the seven unknown parameters $(R_1, k_2, k_{2a}, \gamma, t_D, t_P, \alpha)$, all of which are non-negative. In practice, the investigator may have a rough idea of the range of plausible values for the parameters. In this example, we set the priors as $U(0,20), U(0,10), U(0,10),U(0,5)$  for the first four parameters. For parameters $t_D, t_P, \alpha$, \shortciteN{normandin12} discussed the choice of priors for these parameters and found that the response to a stimulus at 20 minutes should occur before 25 minutes. Here we use for $t_D$ a flat prior around the value 20, so  $t_D \sim U(15, 25)$. This is reasonable to do in most cases because the displacement modelled by $h(t)$ is caused by an external stimulus that the 
experimenter controls and commences at a known time, for example, a drug injection at 20 minutes.
We set the priors for $t_P$ as $U(t_D+1, 35)$ and $\alpha \sim U(0, 25)$; these are essentially the largest numerical ranges that produce sensible simulated data.}

\begin{figure}
\centering
\includegraphics[width=\textwidth, height=120pt]{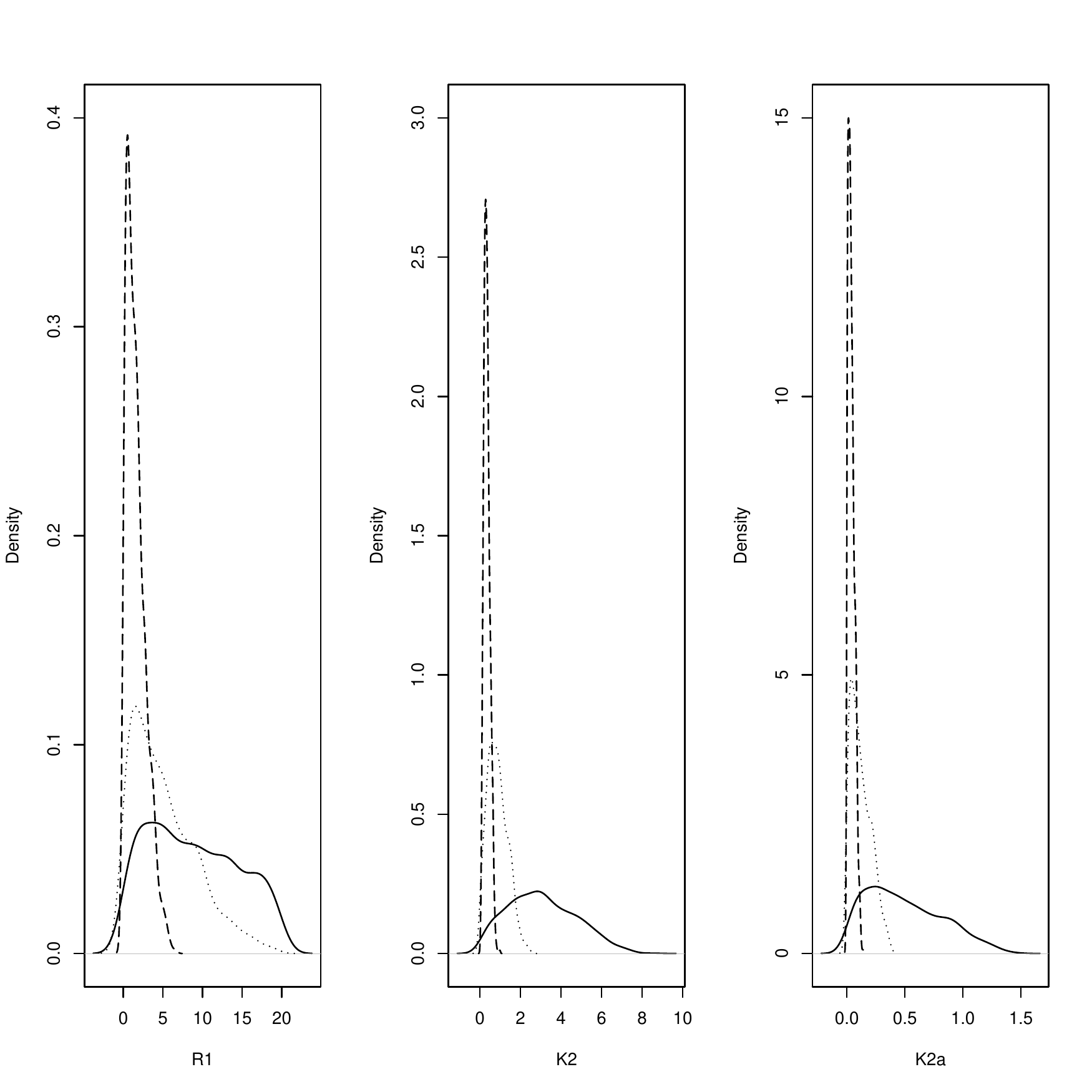}
\caption{\it  Samples for $R_1$, $k_2$ and $k_{2a}$  {after each of the first three iterations,}  indicated by solid,  dotted and dashed lines respectively. Their respective tolerance levels are $\epsilon=200, 50, 10$. }
\label{fig:prior}
\end{figure}

{For an efficient ABC algorithm, we require a good sampling distribution $U(a_i,b_i)$.
 A good starting point for the sampling distribution is to use the prior distributions,  i.e., set $a_i=a_i^*, b_i=b_i^*$. This is typically too diffuse for the algorithm to work efficiently, unless the prior happens to concentrate
around the highest density regions of the posterior. Here we employ a sequential method of narrowing down the range, i.e., finding  values $a_i \geq a_i^*$ 
and $b_i\leq b_i^*$.  We begin by applying the ABC algorithm of  Section \ref{sec:abc}, starting with $a_i = a_i^*$ and $b_i =  b_i^*$, and a large initial tolerance level of $\epsilon=200$. The tolerance is gradually reduced to around 10, over several intermediate steps. With each reduction in the $\epsilon$ value, we use the parameter range obtained from the ABC algorithm at the previous iteration to define new $a_i$ and $b_i$. The samples after each of the first three iterations are plotted in Figure \ref{fig:prior}, for $R_1$, $k_2,$ and $k_{2a}$.  For example, for $k_{2a}$ shown in the right panel,  the first iteration used $U(0,10)$ as the sampling distribution, with a tolerance of $\epsilon=200$. Applying the algorithm of Section \ref{sec:abc}, the range for this parameter has reduced to between 0 and 0.8, as indicated by the solid line. At the next iteration, we use $U(0, 0.8)$
as the new sampling distribution,  with a tolerance of $\epsilon=50$; the dotted line indicates the range for this parameter after the second iteration, which will then
form the sampling distribution for the next iteration, and so on.
The process is then continued until we obtain a reasonably informative range for $U(a_i,b_i)$.  In our simulated dataset, the final sampling distributions were $R_1\sim U(0,5)$, 
$k_2\sim U(0,1)$, $k_{2a} \sim U(0, 0.2)$, $\gamma\sim U(0,2)$. Note that this sequential procedure is valid with the algorithm in Section \ref{sec:abc}, as long as the sampling distribution is proportional to the prior. A more elaborate sequential sampling scheme can be found in \shortciteN{sisson+ft07}.}

\subsection{Summary statistics selection}

We consider four different summary statistics, $S_1,\ldots, S_4$:
\begin{itemize}
\item $S_1$: Spline smoothed data. This is obtained by using the {\tt R} package's {\tt smooth.spline} function, using cross validation. The discrepancy between observed and simulated data is taken as the sum of the absolute differences between the smoothed observed data and the smoothed simulated data {over each time point.}
\item $S_2$: The full data set. The discrepancy between observed and simulated data is taken as the sum of the absolute differences between the raw observed data and the simulated data  {over each time point.}
\item $S_3$: The scaled data set. The discrepancy is the sum of the absolute differences between the raw observed data and the simulated data, where the error at each time point is now scaled by the empirical estimate of the standard deviation of the raw difference.
\item $S_4$: The weighted least squares. For each simulated sample of $t_D, t_P$ and $\alpha$, the weighted least squares estimate of $R_1, k_2, k_{2a}$ and $\gamma$ is estimated for the observed data and simulated data, the discrepancy is taken as the sum of the absolute difference between the four weighed least squares estimates. 
\end{itemize}

\begin{figure}
\centering
\begin{subfigure}
\centering
\includegraphics[width=\textwidth, height=240pt]{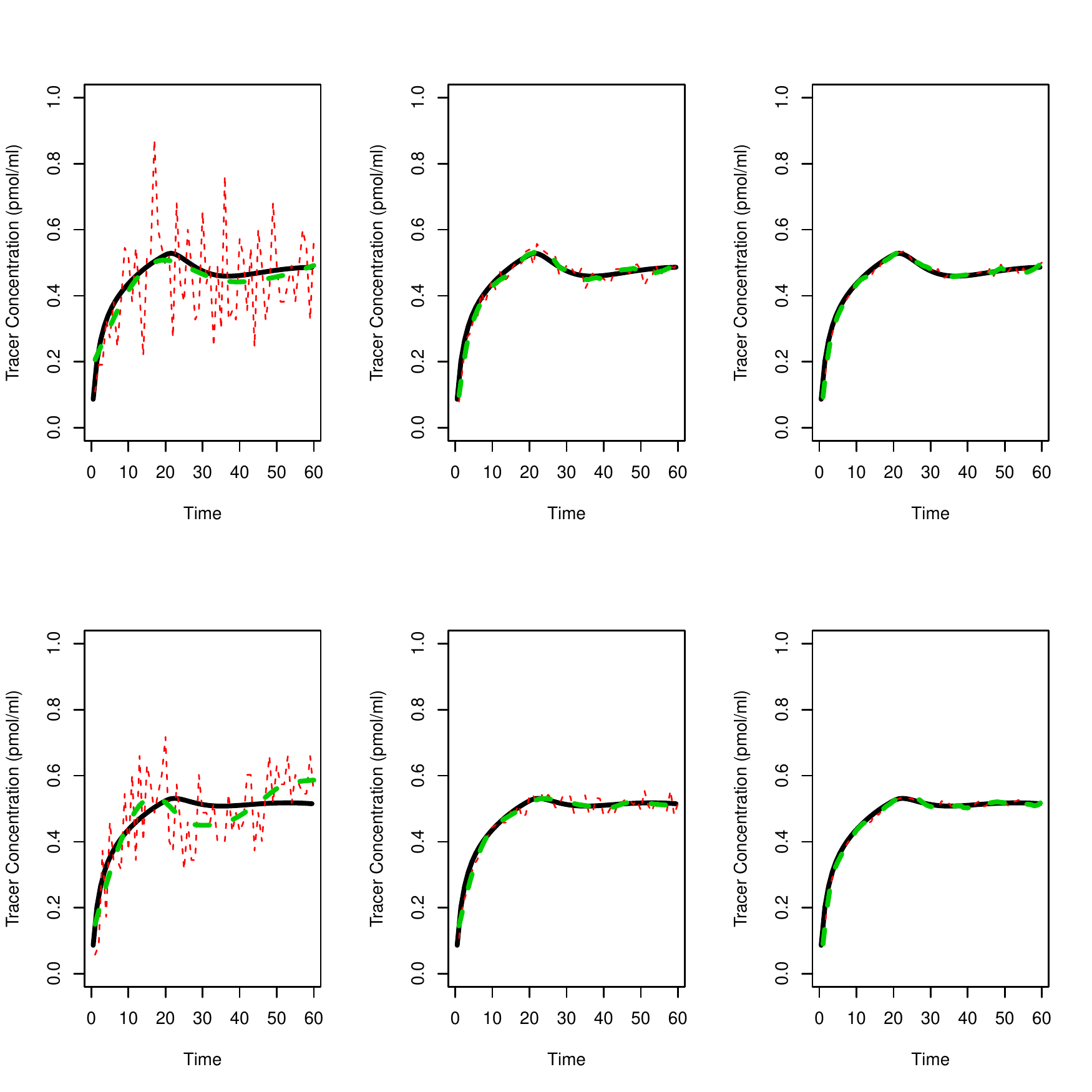}
\end{subfigure}
\begin{subfigure}
\centering
\includegraphics[width=\textwidth, height=240pt]{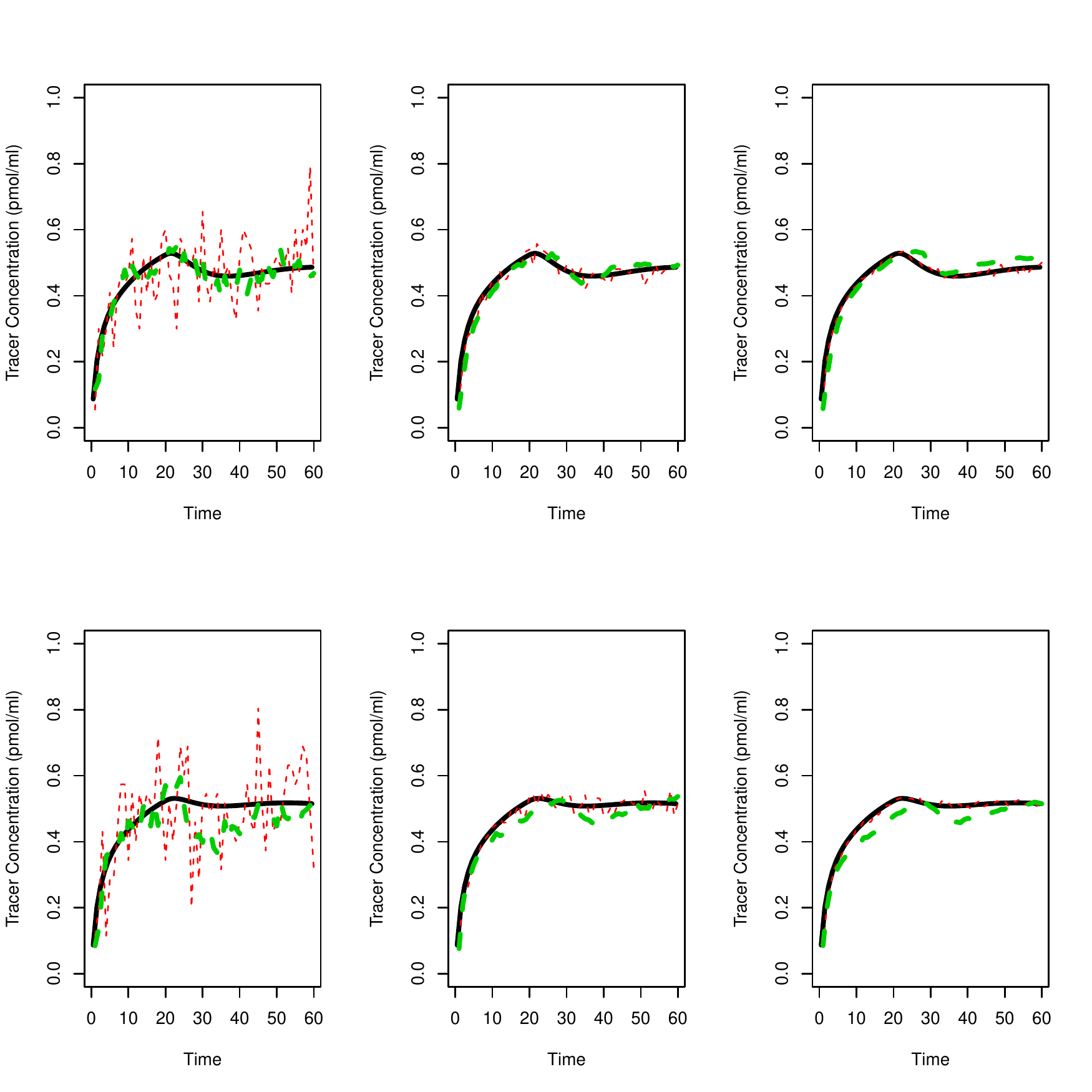}
\end{subfigure}
\caption{\it  {Rows top to bottom correspond to simulated TAC using the model with 200\%  and 100\% activation. Columns from left to right correspond to high to low noise levels.
True mean curves (solid line),  observed noisy data (dotted line),  and smoothed data (dashed line). $S_1$: top two rows. $S_2$: bottom two rows.}}\label{fig:summary12}
\end{figure}

The spline smoothed data can be considered as sample means at each data point, and should be nearly sufficient for the parameters 
of interest. {Figure \ref{fig:summary12} (top two rows) show the TACs for two different activation levels (200\% of baseline activation in the top row and 100\% in the second row, over three different noise levels, ranging from high to low,  shown from left to right). The dotted lines in the figures indicate the raw data, the dashed lines are the spline smoothed estimates of the raw data, and the solid lines are the true (noiseless) curves. 
These {plots} indicate that the spline estimate is very close to the true curve, particularly in low noise level cases, and even in the case of very high noise, it still provides very good estimate of the true TAC.}

Similarly Figure \ref{fig:summary12} (bottom two rows) show the simulated full dataset indicated by dashed lines. The plotted simulated dataset is estimated at a given set of parameter values (not necessarily optimal for the datasets plotted). We can see that at large noise levels, the simulated dataset cannot expect to fully replicate the original dataset, as we do not simulate noise here. Therefore in any ABC applications, when the raw data are used in this way, we do not expect the tolerance to be able to go to zero. In the lower noise levels, the discrepancy between the simulated and observed data is less marked, as would be expected.

\begin{figure}
\centering
\begin{subfigure}
\centering
\includegraphics[width=\textwidth, height=240pt]{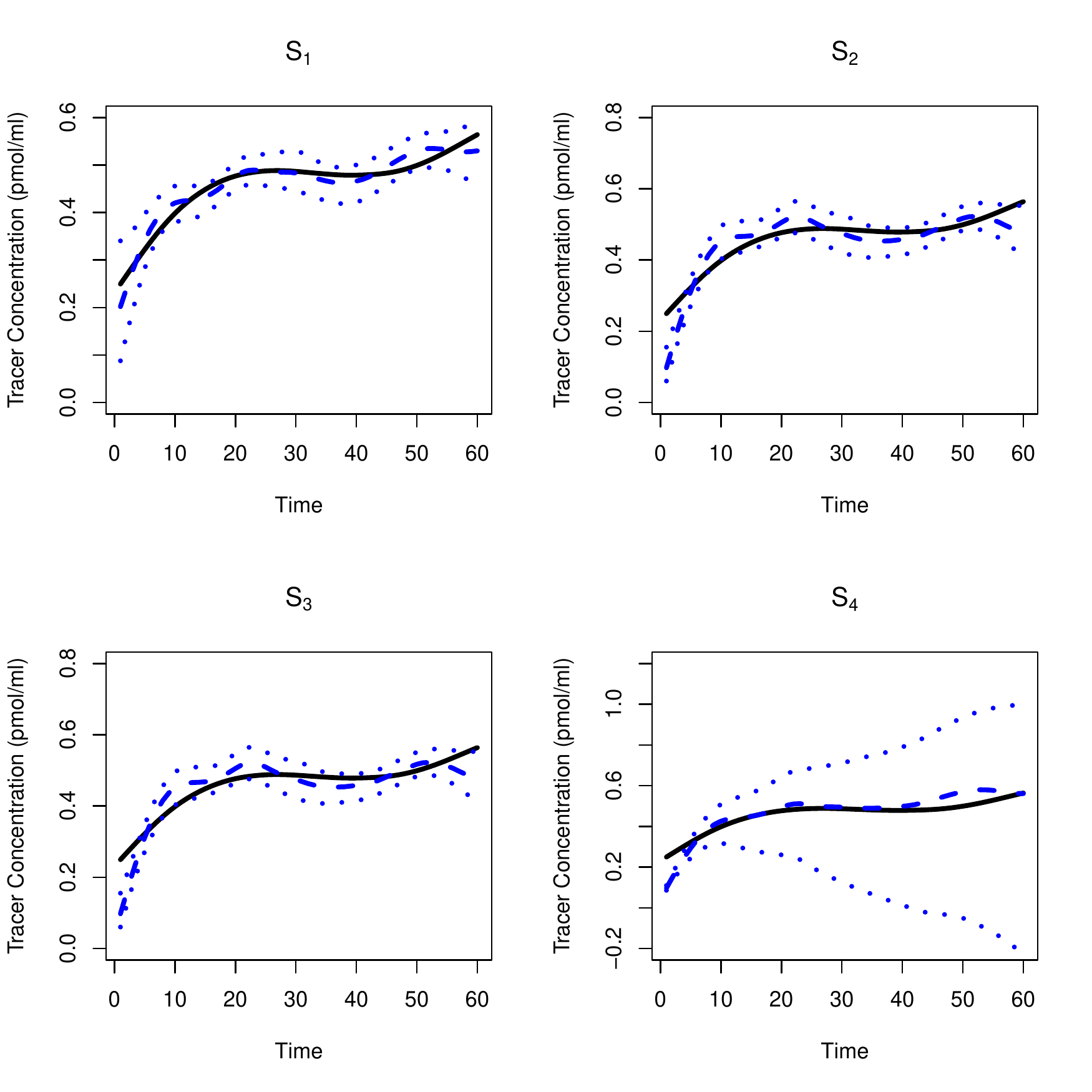}
\end{subfigure}
\begin{subfigure}
\centering
\includegraphics[width=\textwidth, height=240pt]{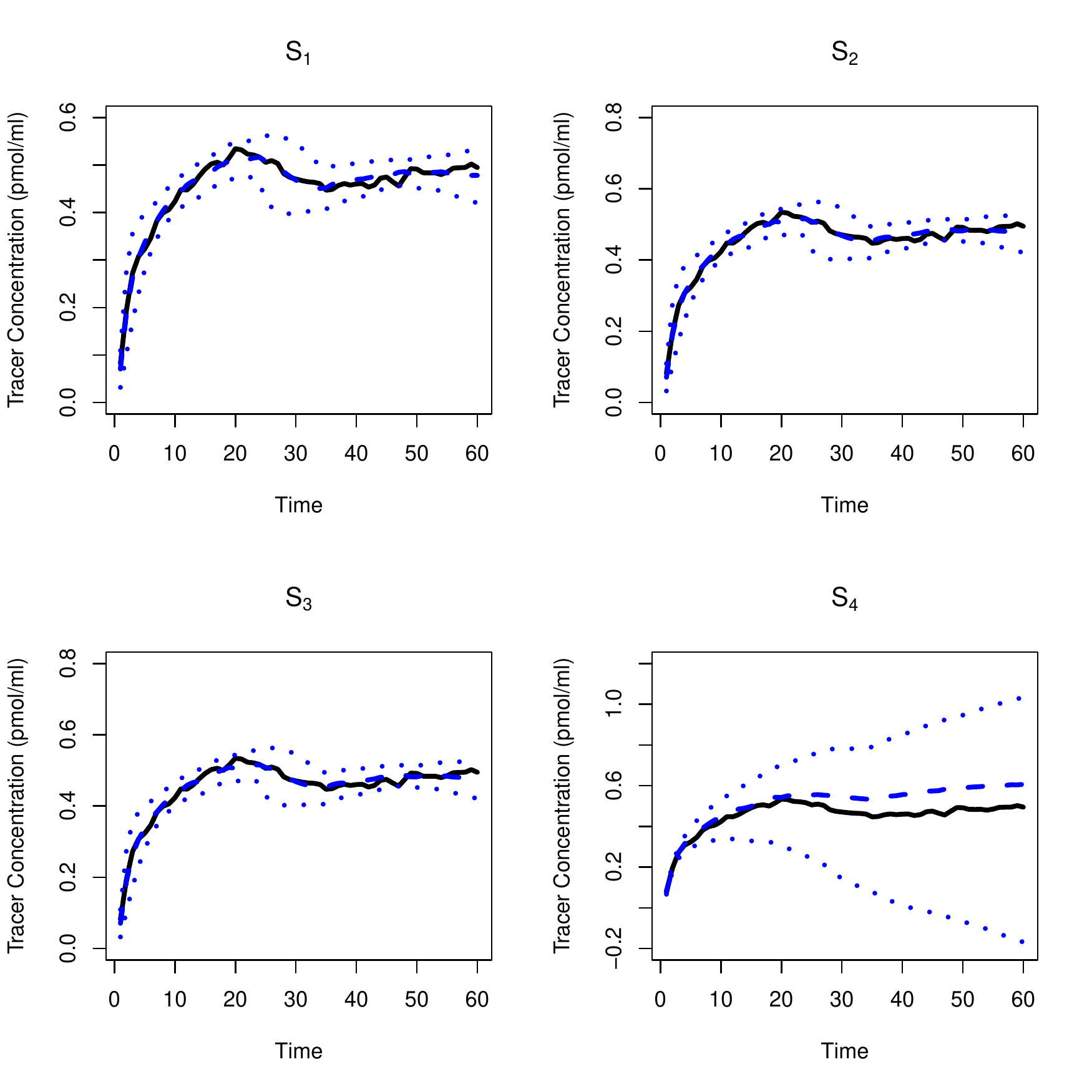}
\end{subfigure}
\caption{\it  {Posterior predictive plots for model with 200\% activation; top two rows at high noise level and bottom two rows at moderate noise level. Results shown for the four different summary statistics $S_1$-$S_4$.  Solid lines indicate smoothed observed data, and dashed and dotted lines are mean, 0.025 and 0.975 percentiles respectively, of the posterior predictive distribution.}}
\label{fig:pp12}
\end{figure}

In order to assess which summary statistics performed best, we considered the use of posterior predictive distributions, \shortciteN{gelman04}.  {For each} summary statistic, we obtain 1 million samples from the sampling distribution and for each statistic, retain the 1000 samples with the smallest error as samples from the posterior. The initial 1 million 
samples {were obtained by setting  $\epsilon = 10$, based on the Euclidean distance between observed and simulated data.} The nominal value of 10 was used because simulation was fast at this value of $\epsilon$, while minimising the burden on computational storage.
For each posterior sample, we generated a dataset and plotted the posterior predictive mean and credibility intervals together with the spline smoothed observed data in 
{Figures \ref{fig:pp12}. The plots show data generated from the model with 200\% activation. Top two rows have a high noise level and the bottom two rows have a moderate noise level.}  Solid lines indicate the observed data, dashed lines indicate the posterior predictive mean, and dotted lines are the corresponding interval limits for a 95\% posterior predictive interval. In both cases, the spline summary $S_1$ performed very well, both in terms of capturing the true curve within the 95\% interval, as well as the fidelity of the estimated curve to the true curve. The full data set, $S_2$ and $S_3$ showed similar performances {to each other}, and gave reasonable performance when the noise level is lower. The weighted least squares estimate $S_4$ performed the worst, and has much more variability in the posterior predictive distribution. In the remainder of this chapter, we will work with $S_1$, the spline smoothed summary.

\subsection{Tolerance level determination}

\begin{figure}
\centering
\includegraphics[width=\textwidth, height=240pt]{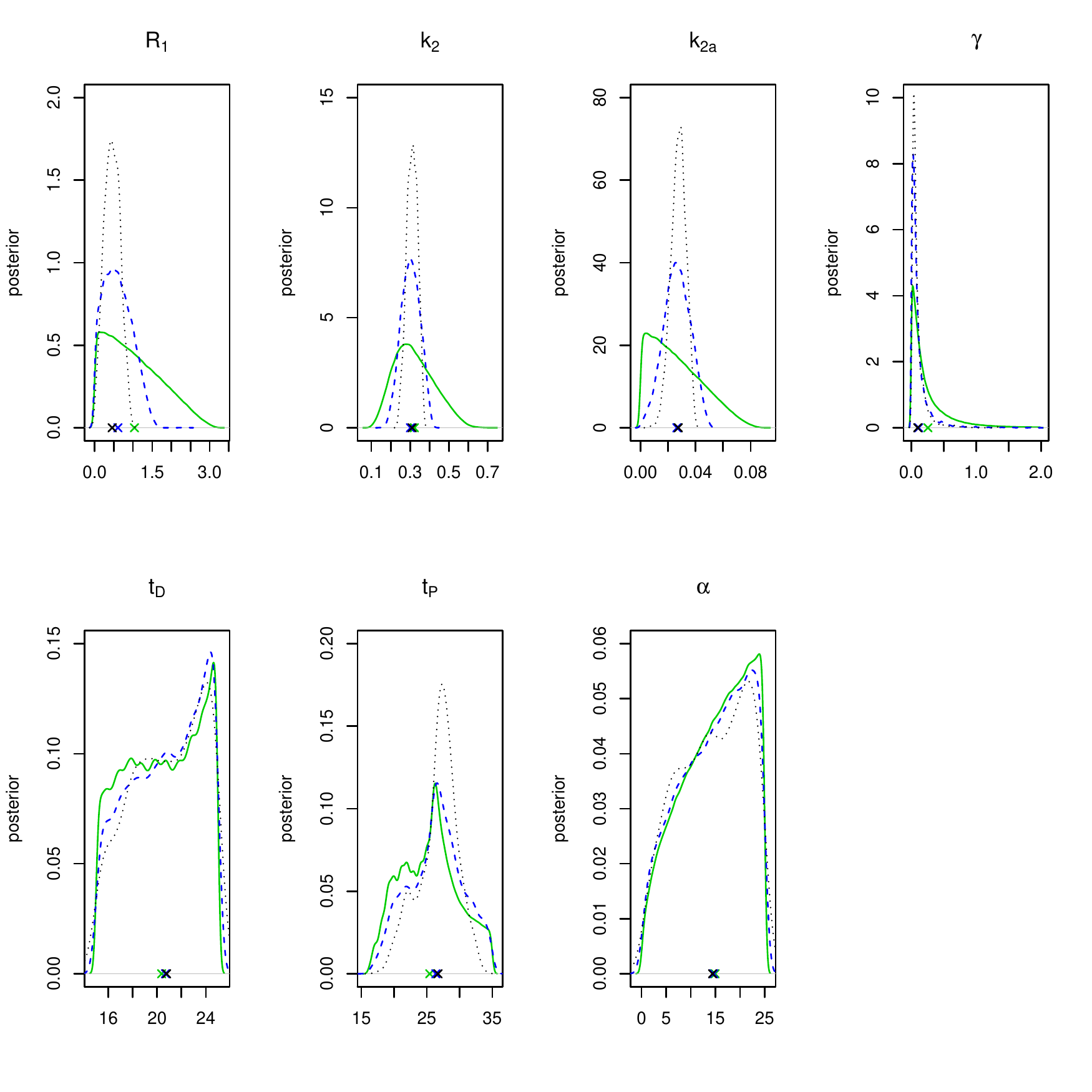}
\caption{\it  Evolution of the {estimated marginal} posterior distribution for the seven parameters, at different values of $\epsilon \approx 7.8, 2.6, 1.7$  {(corresponding to the 0.8, 0.02 and 0.001 percentiles of the one million samples).} Indicated by solid, dashed and dotted lines respectively. $\times$ indicates the posterior means.}
\label{fig:tolerance}
\end{figure}

For the determination of $\epsilon$ in Step 3 of the ABC algorithm, the typical approach is to gradually decrease the value of $\epsilon$ until no further improvements can be made. {Figure \ref{fig:tolerance} illustrates the progression of the {estimated marginal} posteriors at $\epsilon   \approx 7.8, 2.6, 1.7$, corresponding approximately to the 0.8, 0.02 and 0.001 percentiles of the sampled errors in our initial simulation of the one million samples.} The solid line corresponds to the largest error, and {the} dotted line is the one with the smallest error. {Note that the figures show marginal posteriors beyond the range of the prior distributions. This is due to the effect of smoothing for the purpose of visualisation, the true samples should not go beyond the prior distributions.}

 It is evident here that while at larger $\epsilon$ values, the posterior variance is inflated, {the posterior means
do not change too much between varying values of $\epsilon$.}
Interestingly for parameters $\gamma, t_D, t_P$ and $\alpha$, decreasing the values of $\epsilon$ did not produce more information about the parameters, suggesting that the data is fairly uninformative about these parameters. {In our MCMC simulations, we observed similar behaviour with these parameters, suggesting that these parameters of
the lp-ntPET model may not be estimable from the data.}

\subsection{Comparisons of different estimation methods}

\begin{figure}%
\centering
\includegraphics[width=\textwidth, height=240pt]{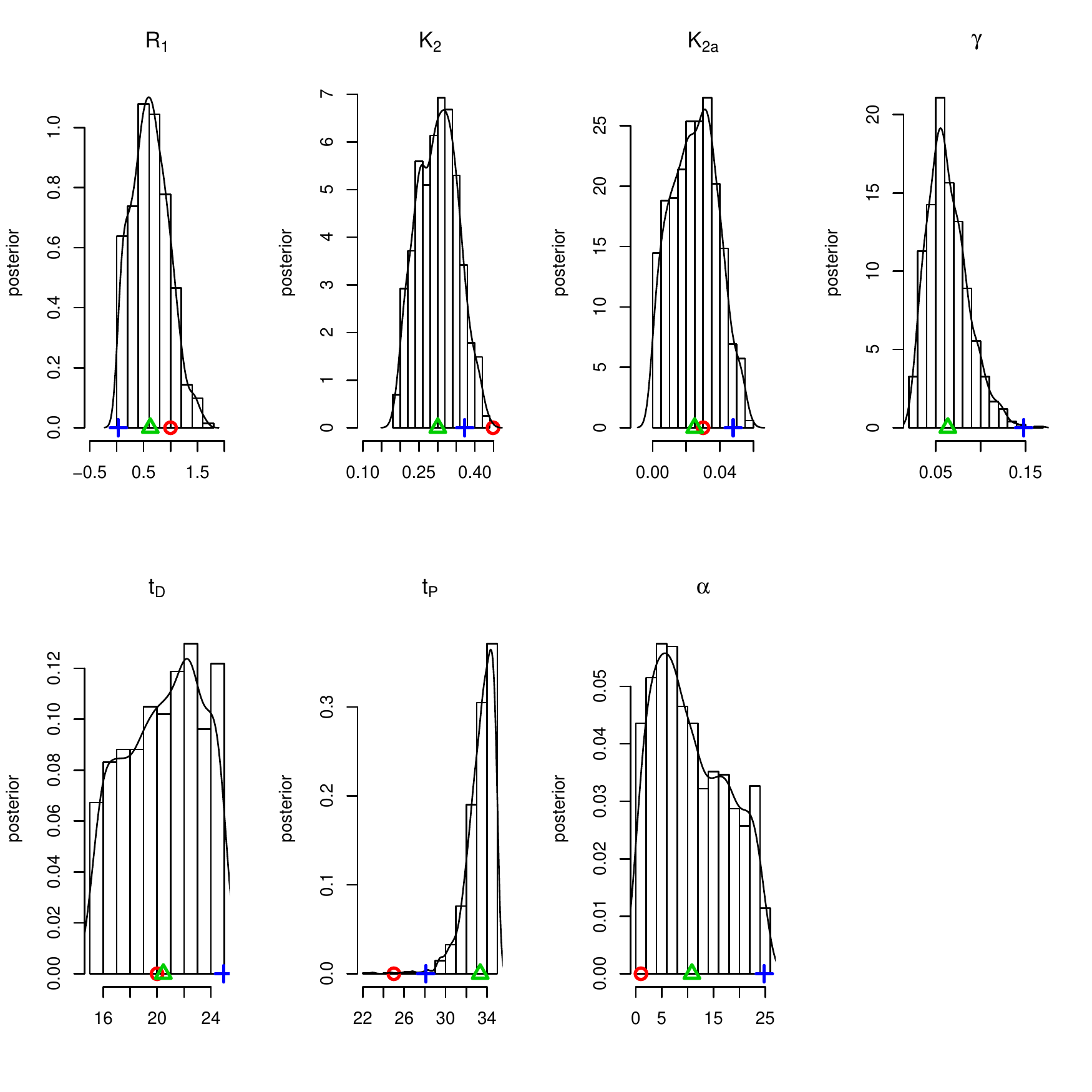}
\caption{\it  {Final estimates of the marginal posterior distributions using ABC, at a value of $\epsilon \approx 1.7$. } Circles indicate the true parameter value, triangles indicate the ABC posterior mean, and plusses indicates the weighted least squares estimate using Normandin et al (2012).} 
\label{fig:post}
\end{figure}

In this section, we compare the performances of ABC, WLS and MCMC on simulation data sets. 
Figure \ref{fig:post} shows the posterior distribution obtained from ABC using the smallest $\epsilon$ value of 1.7, for a single set of simulated data. The model used for the simulation has 200\% activation and a very high noise level.  
Circles indicate the true parameter value used to obtain simulated data, triangles indicate the posterior mean and pluses are the weighted least squares (WLS) estimate of \shortciteN{normandin12}. {For WLS, we }have simulated 100,000 values of $t_D, t_P$ and $\alpha$ from the same prior used for ABC, and computed the estimate following \shortciteN{normandin12}. We have also implemented MCMC assuming {an} independent Gaussian error distribution with variances proportional to the observed TAC. {However, it turns out that the MCMC algorithm is highly sensitive to the starting values, and chains can get stuck easily for many starting points, including those based on the true values.  In addition, the trace plots indicate that the MCMC sampler has bad mixing behaviour, and these appear to be difficult to overcome using the standard MCMC sampler. Most of our MCMC samplers were unable to converge within a reasonable amount of computational time. This may have been caused by the misspecification of the error model, since the errors
in these data are known to be more complicated than Gaussian. The assumption that the variance of the error is proportional to the observed TAC, would likely induce a
highly non-smooth likelihood surface, particularly when data are noisy.
The behaviour of the MCMC output for parameters $t_D$, $t_P$ and $\alpha$ are erratic; these parameters are essentially un-estimable. Indeed, the posterior distribution of these parameters suggest that the data indeed have very little information about the values of $t_D$ and $\alpha$, as the posteriors are largely unchanged from our prior distribution. This is also seen in the results from ABC,  shown in Figure \ref{fig:post}. }
We found that MCMC tended to over-estimate the $R_1$ parameter while this is underestimated by ABC and WLS in some cases. MCMC was able to give very precise estimates of $t_D$ close to the true value, while it found it difficult to estimate $t_P$. The situation is reversed for ABC, which found $t_D$ difficult to estimate while $t_P$ was relatively straight-forward. It is difficult to know the exact reason for these discrepancies, a lack of convergence in the MCMC sampler could partially explain some of the differences, model misspecification is another possibility.

\begin{figure}%
\centering
\includegraphics[width=\textwidth, height=240pt]{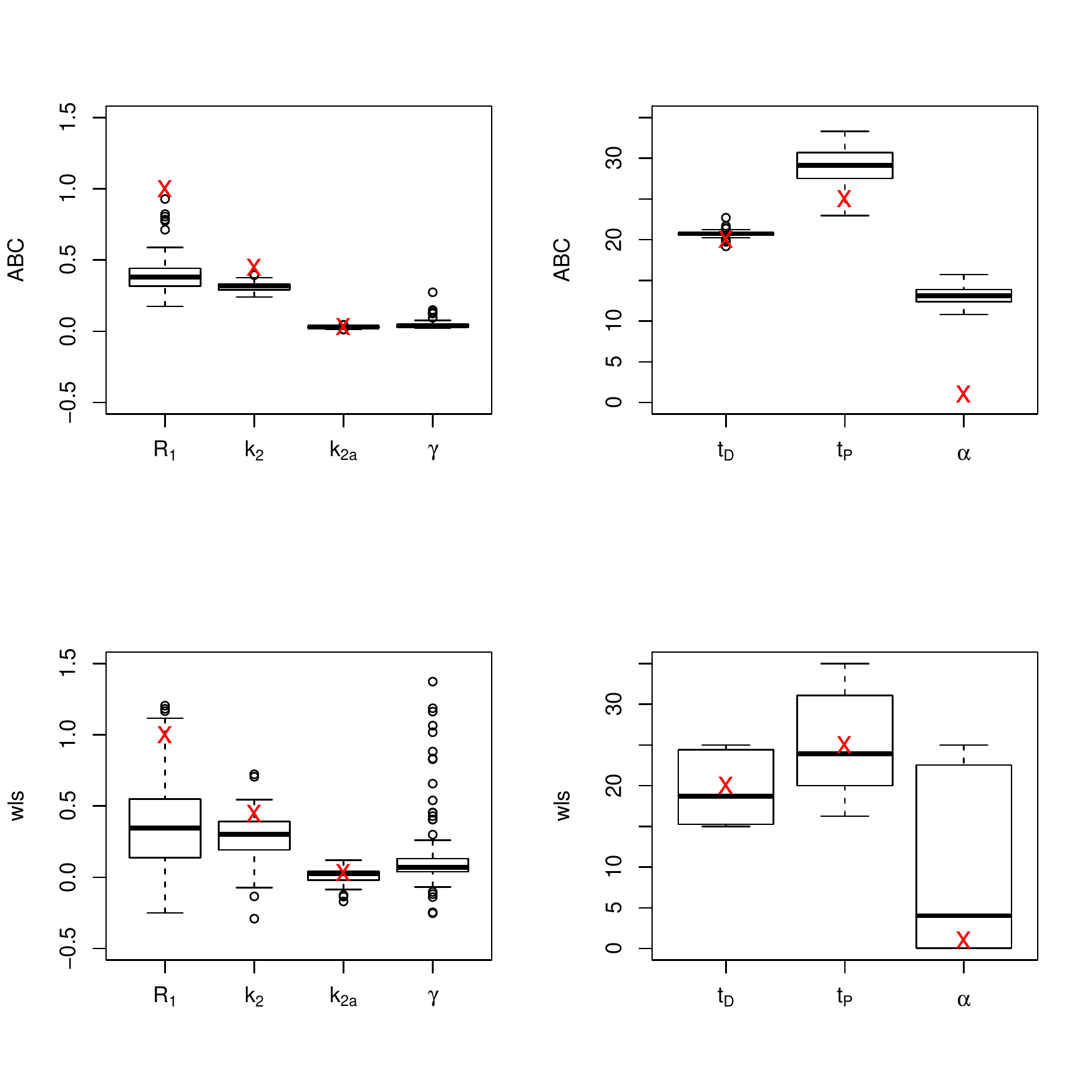}
\caption{\it  Boxplots of posterior mean estimates over 100 noise realisations using ABC (top row), the corresponding WLS estimates (bottom row).  Crosses indicate the true values.}
\label{fig:boxplots}
\end{figure}

Figure \ref{fig:boxplots} shows the posterior mean estimates from ABC (top row) and least squares estimates for 100 noise realisations (at 100\% activation and highest noise level). We have excluded results from MCMC simulations due to the unreliable results obtained. At these noise and activation specifications, the parameter estimations were the most problematic. Results in Figure \ref{fig:boxplots} demonstrate that {ABC estimates are much less variable than WLS}, although for both algorithms, the parameter $R_1$ is largely underestimated. The ABC estimator is more robust for all the parameters, but particularly so for $R_1$ and the time course response parameters $t_D$, $t_P$ and $\alpha$, where the variability of the estimates as demonstrated by the box plots are much smaller.

\begin{figure}%
\centering

\centering
\includegraphics[width=5.5cm, height=240pt]{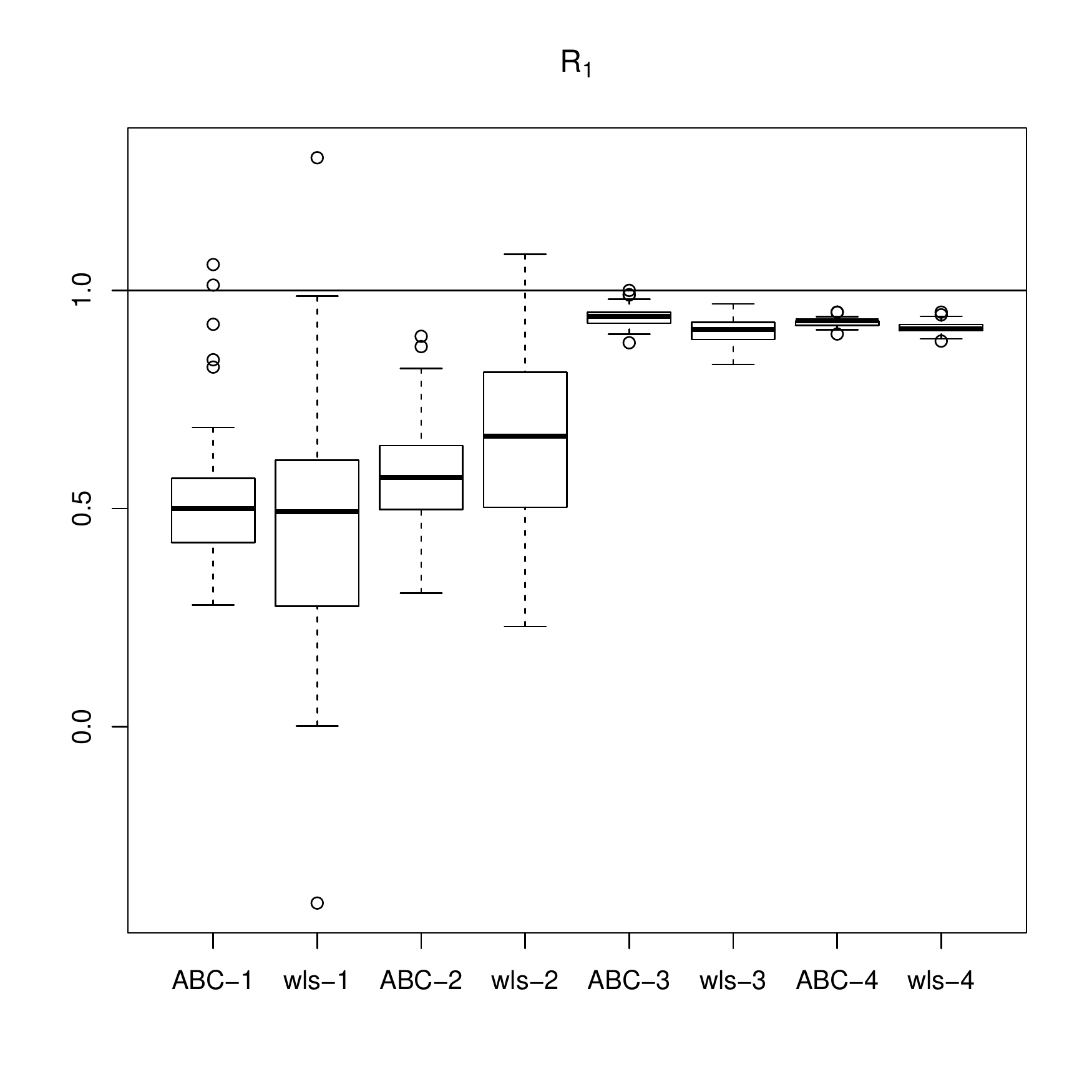}
\centering
\includegraphics[width=5.5cm, height=240pt]{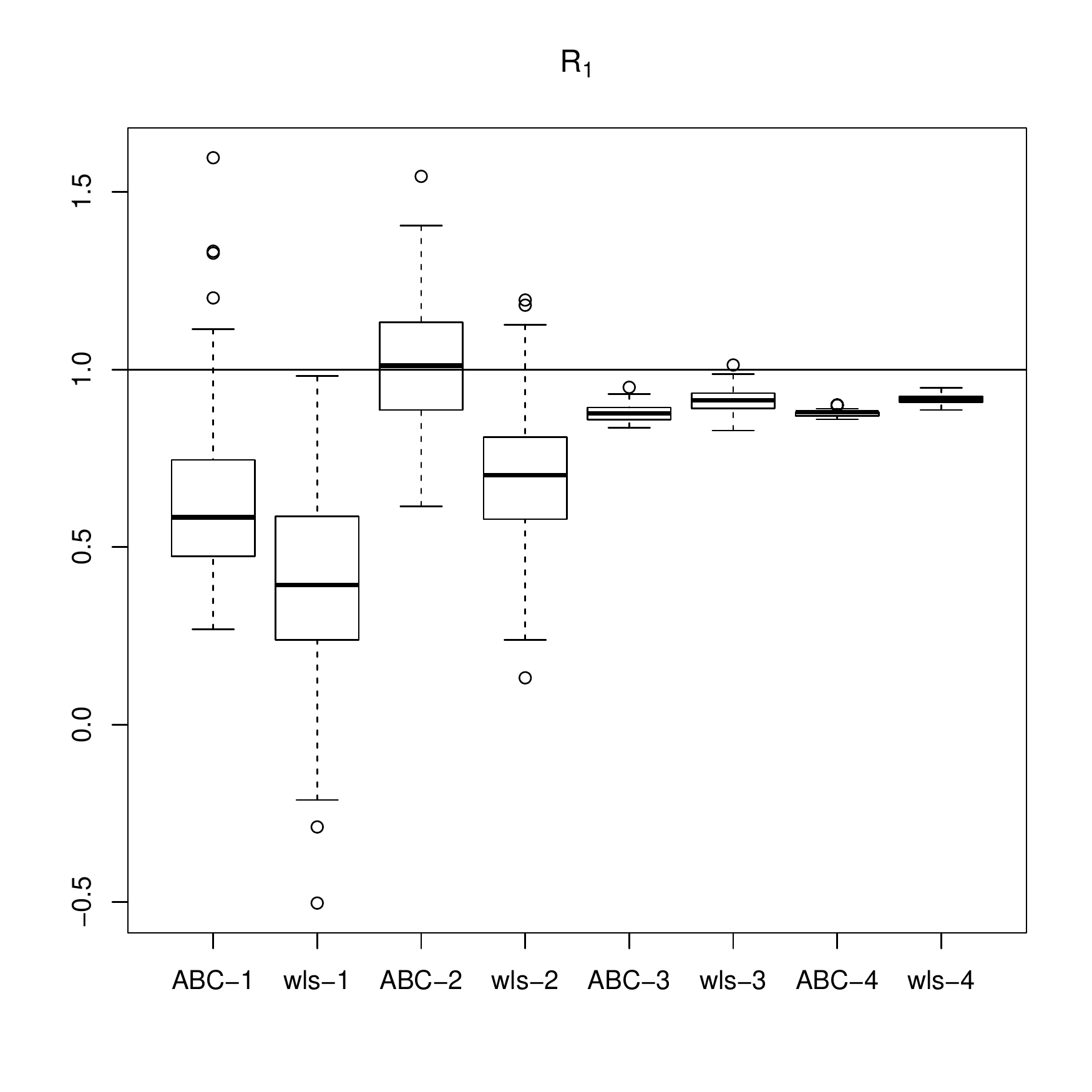}
\caption{\it  Boxplots of posterior mean estimates over 100 noise realisations based on 100\% activation model (left) and 200\% activation model (right). {X-axis correspond to  results} from ABC and WLS alternately, for four different noise levels (1-4) from highest to lowest noise levels. Horizontal line indicates the true value at 1.}
\label{fig:boxplotsR1}
\end{figure}

We further investigated the cause for the apparently large bias in the $R_1$ estimation in both ABC and WLS. We found that parameter estimates are somewhat sensitive to the prior specification of the parameter $\alpha$, and using a smaller range of $U(0,3)$ we were able to obtain better estimates for both algorithms. However, this {still} did not provide a substantial  improvement to the {bias in the }$R_1$ estimates. Figure \ref{fig:boxplotsR1} shows the comparative box plots for the $R_1$ parameter as estimated by ABC and WLS, over four noise levels and two different activation levels (200\% and 100\%). While it can be seen that the estimates of the 200\% activation model are generally better than the 100\% activation model, in both cases, the estimates worsen with noise, exhibiting high bias and high variance. The performance of the ABC estimator in the higher noise cases are generally superior to WLS. In low noise cases, WLS are often similar or even better than ABC, suggesting that the benefit of a Bayesian analysis lies in the more noisy problems.

In terms of the large bias in $R_1$, one possibility is that it could be an inherent bias of the lp-ntPET model, but this seems unlikely to explain {away} all the bias. Figure \ref{fig:boxplotsR1} suggests that for lower noise levels, the results are close to the true value. This suggests that the biases maybe due to the way we handle {the} noise. A closer look at Figures \ref{fig:summary12}, {first column of top two rows}, suggests that the spline based summary deviates from the true TAC, while in lower noise data, there is much better accordance between the summaries and the true curve. This suggests that a more robust spline estimator, less sensitive to the distribution of the noise,  may yield better results. 

\section{Conclusions and Discussions}\label{sec:discuss}
This chapter {examined} the use of ABC for medical imaging data. In these types of data, it is often necessary to perform parameter estimation for multiple datasets, sometimes in the order of tens of thousands. A computational advantage of ABC in this scenario is that simulation of synthetic datasets within the ABC step will only need to be done once, representing a substantial computational saving
compared with more traditional estimation procedures such as MCMC.

Our simulation studies comparing ABC, MCMC and WLS showed that MCMC was unstable, and difficult to implement {under our model assumptions}.  ABC and WLS obtained comparable results, and in most cases, were able to retrieve the true parameter values. In higher noise problems, ABC produced more robust estimates than WLS, which will prove more useful for voxel-wise estimations. {In terms of computational time, WLS is the fastest. Both ABC and MCMC are time consuming, but over multiple datasets, ABC is substantially faster than MCMC.}

We expect that in less noisy datasets, with relatively simple kinetic models, WLS would perform well, and it would be difficult to justify the use of the more computationally expensive ABC method. However, even in this case, there are added benefits from a Bayesian analysis that are often not readily available from the frequentist approach. {For instance, \shortciteN{normandin12} were interested in
the significance of the magnitude parameter $\gamma$. However finding an appropriate statistical test for such a task is difficult. In Bayesian inference, the posterior distribution of $\gamma$ from Figure \ref{fig:post} readily provides the credibility interval for the parameter, and allows us to assess the significance of a parameter immediately.}  Alternatively, posterior model comparison can be carried out relatively straight-forwardly; see elsewhere in this volume for more details on ABC model choice.  Finally, the posterior distribution provides some information on how well the data are able to estimate certain parameters in a given model, see for example parameters $t_D$ and $\alpha$, and this could serve as an exploratory tool for the development of new models.

The ABC algorithm described in this chapter {shares some similarity to} the WLS approach of  \shortciteN{normandin12}, where a basis function approach is used to estimate the time course response curve. Their WLS can be seen as a hybrid of Bayesian and frequentist methods. The main differences are that ABC requires the selection of a summary statistic and WLS searches for the modal estimate while ABC computes the full posterior. In both algorithms, parameters become harder to estimate when the noise level is high. One possibility in ABC is to consider summaries which are more robust to noise. Another possible direction is to extend the analysis, currently assuming voxel independence, to allow borrowing of information from nearby voxels. It would be interesting to see how this can be performed {efficiently} within the ABC setting.

\bibliographystyle{chicago}
\bibliography{ABCPET}

\begin{thebibliography}{}

\bibitem[\protect\citeauthoryear{Alpert and Yuan}{Alpert and
  Yuan}{2009}]{alpert09}
Alpert, N.~M. and F.~Yuan (2009).
\newblock A general method of {B}ayesian estimation for parametric imaging of
  the brain.
\newblock {\em Neuroimage\/}~{\em 45}, 1183--1189.

\bibitem[\protect\citeauthoryear{Anger}{Anger}{1964}]{anger64}
Anger, H.~O. (1964).
\newblock Scintillation camera with multichannel collimators.
\newblock {\em Journal of Nuclear Medicine\/}~{\em 65}, 515--531.

\bibitem[\protect\citeauthoryear{Carson}{Carson}{1986}]{carson86}
Carson, R.~E. (1986).
\newblock {\em Positron emission tomography and autoradiography: Principles and
  applications for the Positron emission tomography and autoradiography:
  Principles and applications for the brain and heart}, Chapter Parameter
  estimation in positron emission tomography, pp.\  347-- 390.
\newblock Raven Press, New York.

\bibitem[\protect\citeauthoryear{Carson, Huang, and Green}{Carson
  et~al.}{1986}]{carsonh86}
Carson, R.~E., S.~C. Huang, and M.~E. Green (1986).
\newblock Weighted integration method for local cerebral blood flow
  measurements with positron emission tomography.
\newblock {\em Journal of Cerebral Blood Flow and Metabolism\/}~{\em 6},
  245--258.

\bibitem[\protect\citeauthoryear{Cherry and Dahlbom}{Cherry and
  Dahlbom}{2004}]{cherry04}
Cherry, S.~R. and M.~Dahlbom (2004).
\newblock {\em {PET}: molecular imaging and its biological applications},
  Chapter {PET}: physics, instrumentation and scanners, pp.\  1--124.
\newblock Berlin: Springer.

\bibitem[\protect\citeauthoryear{Cunningham and Jones}{Cunningham and
  Jones}{1993}]{cunninghamj93}
Cunningham, V.~J. and T.~Jones (1993).
\newblock Spectral analysis of dynamic {PET} studies.
\newblock {\em Journal of Cerebral Blood Flow and Metabolism\/}~{\em 13\/}(1),
  15--23.

\bibitem[\protect\citeauthoryear{Feng, Huang, Wang, and Ho}{Feng
  et~al.}{1996}]{feng96}
Feng, D., S.~C. Huang, Z.~Wang, and D.~Ho (1996).
\newblock An unbiased parametric imaging algorithm for uniformly sampled
  biomedical system parameter estimation.
\newblock {\em {IEEE} {T}ransactions on Medical Imaging\/}~{\em 15}, 521--518.

\bibitem[\protect\citeauthoryear{Fessler}{Fessler}{1996}]{fessler96}
Fessler, J.~A. (1996).
\newblock Mean and variance of implicitly defined biased estimators (such as
  penalized maximum likelihood): applications to tomography.
\newblock {\em IEEE Transactions on Image Processing\/}~{\em 5}, 493--506.

\bibitem[\protect\citeauthoryear{Garthwaite, Fan, and Sisson}{Garthwaite
  et~al.}{2015}]{garthwaitefs15}
Garthwaite, P.~H., Y.~Fan, and S.~A. Sisson (2015).
\newblock Adaptive optimal scaling of metropolis-hastings algorithms using the
  robbins-monro process.
\newblock {\em Communications in Statistics - Theory and Methods\/}, In Press.

\bibitem[\protect\citeauthoryear{Gelman, Carlin, Stern, and Rubin}{Gelman
  et~al.}{2004}]{gelman04}
Gelman, A., J.~B. Carlin, H.~S. Stern, and D.~B. Rubin (2004).
\newblock {\em Bayesian Data Analysis}.
\newblock Texts in Statistical Science. Chapman and Hall/CRC Press.

\bibitem[\protect\citeauthoryear{Gunn, Gunn, Turkheimer, Aston, and
  Cunningham}{Gunn et~al.}{2002}]{gunngtac02}
Gunn, R.~N., S.~R. Gunn, F.~E. Turkheimer, J.~A.~D. Aston, and V.~J. Cunningham
  (2002).
\newblock Positron {E}mission {T}omography compartmental models; a basis
  pursuit strategy for kinetic modeling.
\newblock {\em Journal of Cerebral Blood Flow and Metabolism\/}~{\em 22},
  1425--1439.

\bibitem[\protect\citeauthoryear{Gunn, Lammertsma, Hume, and Cunningham}{Gunn
  et~al.}{1997}]{gunn97}
Gunn, R.~N., A.~A. Lammertsma, S.~P. Hume, and V.~J. Cunningham (1997).
\newblock Parametric imaging of ligand-receptor binding in {PET} using a
  simplified reference region model.
\newblock {\em Neroimage\/}~{\em 6\/}(4), 279--287.

\bibitem[\protect\citeauthoryear{Gunn, Slifstein, Searle, and Price}{Gunn
  et~al.}{2015}]{gunnssp15}
Gunn, R.~N., M.~Slifstein, G.~E. Searle, and J.~C. Price (2015).
\newblock Quantitative imaging of protein targets in the human brain with
  {PET}.
\newblock {\em Physics in Medicine and Biology\/}~{\em 60}, R363--R411.

\bibitem[\protect\citeauthoryear{Hudson and Larkin}{Hudson and
  Larkin}{1994}]{hudsonl94}
Hudson, H.~M. and R.~S. Larkin (1994).
\newblock Accelerated image reconstruction using ordered subsets of projection
  data.
\newblock {\em IEEE Transactions on medical imaging\/}~{\em 13\/}(4), 601--609.

\bibitem[\protect\citeauthoryear{Innis and {et al}}{Innis and {et
  al}}{2007}]{innis07}
Innis, R.~B. and {et al} (2007).
\newblock Consensus nomenclature for in vivo imaging of reversibly binding
  radioligands.
\newblock {\em Journal of Cerebral Blood Flow and Metabolism\/}~{\em 27},
  1533--1539.

\bibitem[\protect\citeauthoryear{Leahy and Qi}{Leahy and Qi}{2000}]{leahyq00}
Leahy, R.~M. and J.~Qi (2000).
\newblock Statistical approaches in quantitative positron emission tomography.
\newblock {\em Statistics and Computing\/}~{\em 10}, 147 -- 165.

\bibitem[\protect\citeauthoryear{Lin, Haldar, Li, Conti, and Leahy}{Lin
  et~al.}{2014}]{lin2014}
Lin, Y., J.~Haldar, Q.~Li, P.~Conti, and R.~Leahy (2014).
\newblock Sparsity constrained mixture modeling for the estimation of kinetic
  parameters in dynamic {PET}.
\newblock {\em {IEEE} Transactions on Medical Imaging\/}~{\em 33\/}(173-185).

\bibitem[\protect\citeauthoryear{Malave and Sitek}{Malave and
  Sitek}{2015}]{malaves15}
Malave, P. and A.~Sitek (2015).
\newblock Bayesian analysis of a one-compartment kinetic model used in medical
  imaging.
\newblock {\em Journal of Applied Statistics\/}~{\em 42\/}(1), 98--113.

\bibitem[\protect\citeauthoryear{Meikle, Matthews, Cunningham, Bailey,
  Livieratos, Jones, and Price}{Meikle et~al.}{1998}]{meikle98}
Meikle, S.~R., J.~C. Matthews, V.~J. Cunningham, D.~L. Bailey, L.~Livieratos,
  T.~Jones, and P.~Price (1998).
\newblock Parametric image reconstruction using spectral analysis of {PET}
  projection data.
\newblock {\em Physics in Medicine and Biology\/}~{\em 43}, 651--666.

\bibitem[\protect\citeauthoryear{Morris, Yoder, Wang, Normandin, Zheng, Mock,
  Jr, and Froehlich}{Morris et~al.}{2005}]{morris05}
Morris, E.~D., K.~K. Yoder, C.~Wang, M.~Normandin, Q.-H. Zheng, B.~Mock,
  R.~F.~M. Jr, and J.~C. Froehlich (2005).
\newblock {ntPET}: A new application of {PET} imaging for characterizing the
  kinetics of endogenous neurotransmitter release.
\newblock {\em Molevular Imaging\/}~{\em 4\/}(4), 473--489.

\bibitem[\protect\citeauthoryear{Morris, Enders, Schmidt, Bradley, Jr, and
  Fisher}{Morris et~al.}{2004}]{morris04}
Morris, E.~V., C.~J. Enders, K.~C. Schmidt, T.~C. Bradley, R.~F.~M. Jr, and
  E.~E. Fisher (2004).
\newblock {\em Emission Tomography: The Fundamentals of {PET} and {SPECT}},
  Chapter Kinetic modeling in positron emission tomography, pp.\  499--540.
\newblock Academic Press, Amsterdam; Boston.

\bibitem[\protect\citeauthoryear{Normandin, Schiffer, and Morris}{Normandin
  et~al.}{2012}]{normandin12}
Normandin, M.~D., W.~K. Schiffer, and E.~D. Morris (2012).
\newblock A linear model for estimation of neurotransmitter response profiles
  from dynamic {PET} data.
\newblock {\em Neuroimage\/}~{\em 59}, 2689--2699.

\bibitem[\protect\citeauthoryear{Qi and Leahy}{Qi and Leahy}{2006}]{qil06}
Qi, J. and R.~M. Leahy (2006).
\newblock Iterative reconstruction techniques in emission computed tomography.
\newblock {\em Physics in Medicine and Biology\/}~{\em 51\/}(541-578).

\bibitem[\protect\citeauthoryear{Shepp and Vardi}{Shepp and
  Vardi}{1982}]{sheppv82}
Shepp, L.~A. and Y.~Vardi (1982).
\newblock Maximum likelihood reconstruction for emission tomography.
\newblock {\em IEEE Transactions on medical imaging\/}~{\em MI-1\/}(2),
  113--122.

\bibitem[\protect\citeauthoryear{Sisson, Fan, and Tanaka}{Sisson
  et~al.}{2007}]{sisson+ft07}
Sisson, S.~A., Y.~Fan, and M.~M. Tanaka (2007).
\newblock {Sequential Monte Carlo} without likelihoods.
\newblock {\em Proceedings of the National Academy of Sciences\/}~{\em 104},
  1760--1765. Errata (2009), 106:16889.

\bibitem[\protect\citeauthoryear{Sitek}{Sitek}{2014}]{sitek14}
Sitek, A. (2014).
\newblock {\em Statistical Computing in Nuclear Imaging}.
\newblock Series in Medical Physics and Biomedical Engineering. CRC Press.

\bibitem[\protect\citeauthoryear{Wernick and Aarsvold}{Wernick and
  Aarsvold}{2004}]{wernick04}
Wernick, M.~N. and J.~N. Aarsvold (2004).
\newblock {\em Emission tomography: The fundamentals of {PET} and {SPECT}}.
\newblock New York: Academic.

\bibitem[\protect\citeauthoryear{Zhou, Aston, and Johansen}{Zhou
  et~al.}{2013}]{zhouaj13}
Zhou, Y., J.~A.~D. Aston, and A.~M. Johansen (2013).
\newblock Bayesian model comparison for compartmental models with applications
  in positron emission tomography.
\newblock {\em Journal of Applied Statistics\/}~{\em 40}, 993--1016.

\bibitem[\protect\citeauthoryear{Zhou, Huang, and Bergsneider}{Zhou
  et~al.}{2001}]{zhou01}
Zhou, Y., S.~C. Huang, and M.~Bergsneider (2001).
\newblock Linear ridge regression with spatial constraint for generation of
  parameter images in dynamic positron emission tomography studies.
\newblock {\em IEEE Transactions on Nuclear Science\/}~{\em 48}, 125--130.

\end{thebibliography}


\end{document}